\documentclass[aps,prl,twocolumn,groupedaddress]{revtex4}
\usepackage{graphicx}

\usepackage{dcolumn}

\usepackage{bm}

\usepackage{color}

\begin{document}

\title{Surface Charge Density Wave Transition in NbSe$_3$}

\author{Christophe Brun$^{1,*}$, Zhao-Zhong Wang$^1$, Pierre Monceau$^2$, and Serguei Brazovskii$^3$}

\affiliation{$^1$ Laboratoire de Photonique et de Nanostructures, CNRS, route de Nozay, 91460 Marcoussis, France\\$^2$ Institut N\'eel, CNRS and University Joseph Fourier, 25 Avenue des Martyrs, B.P. 166, 38042 Grenoble cedex 9, France\\$^3$ Laboratoire de Physique Th\'eorique et Mod\`eles Statistiques, CNRS and University Paris-Sud, bat. 100, 91405  Orsay, France\\}

\date{13-01-2010}

\begin{abstract}
The two charge-density wave (CDW) transitions in NbSe$_3$
were investigated by scanning tunneling microscopy (STM) on \emph{in situ} cleaved $(\bm{b},\bm{c})$ plane. The temperature dependence of first-order CDW satellite spots, obtained from the Fourier transform of the STM images, was measured between 5-140 K to extract the surface critical temperatures (T$_s$). The low T CDW transition occurs at T$_{2s}$=70-75 K, more than 15 K above the bulk T$_{2b}=59$K while at exactly the same wave number.
Plausible mechanism for such an unusually high surface enhancement is a softening of transverse phonon modes involved in the CDW formation.
The regime of 2D fluctuations is analyzed according to a Berezinskii-Kosterlitz-Thouless type of surface transition, expected for this incommensurate 2D CDW, by extracting the temperature dependence of the order parameter correlation functions.
\end{abstract}


\maketitle

Powerful techniques have been recently developed, such as grazing incidence inelastic x-ray scattering, angle resolved photoemission, or scanning tunneling microscopy (STM), to probe the surface layer(s) of correlated electron systems. An important question is currently raised regarding the surface electronic states of these systems: are they identical to the bulk ones? This question becomes even more crucial when the material undergoes a phase transition to a broken symmetry state, for instance a charge-density wave (CDW) transition; the free surface must reflect this broken symmetry. The "extraordinary transition" corresponds to the very interesting case where the surface orders at higher temperature than the bulk \cite{Binder1983}. It opens the opportunity to study a truly two-dimensional (2D) system which is more perfect than usual ones prepared on a substrate.

Up to now experimental systems known to reveal such surface transitions are extremely scarce \cite{Pleimling2004}. The antiferromagnetic system NiO(100) is one of these \cite{Marynowski1999}. Recently, at the surface of the quasi-2D compound NbSe$_2$, the CDW was shown to order at about $1.5$ K above the bulk transition temperature \cite{Murphy2003}. This was subsequently corroborated by emphasizing a modified behavior of the surface phonon modes \cite{Murphy2005}. On the other hand "the commensurate checkerboard order" discovered by STM in the cuprate system Ca$_{2-x}$Na$_x$CuO$_2$Cl$_{2}$ (NaCCOC) \cite{Hanaguri2004}, furnishes a example of charge ordering stabilized at the surface whereas the bulk doesn't show such counterpart at lower temperature. Hereafter, we report high-resolution \emph{in situ} STM measurements on the quasi-one dimensional (1D) compound NbSe$_3$ which undergoes in the bulk two independent CDW transitions at $T_{1b}=144$ K and at $T_{2b}=59$ K, involving two CDW vectors, $\bm{q_1} = 0.24$ $\bm{b^*}$ for the high-temperature (HT) CDW and $\bm{q_2}= 0.5$ $\bm{a^*} + 0.26$ $\bm{b^*} + 0.5$ $\bm{c^*}$ for the low-temperature (LT) CDW. By measuring the temperature dependence of the amplitude of the $\bm{q_1}$ and $\bm{q_2}$ first-order satellites we found that at the $(\bm{b},\bm{c})$ surface of NbSe$_3$, the $\bm{q_2}$ transition occurs at 15 K higher than in the bulk, demonstrating the huge effect of the surface on the LT CDW phase transition. Owing to the phase degeneracy of this incommensurate CDW and to the anisotropic electronic properties of NbSe$_3$, the Berezinskii-Kosterlitz-Thouless (BKT) transition is investigated for the first time in real space.

\begin{figure*}
\includegraphics[width=13.5cm]{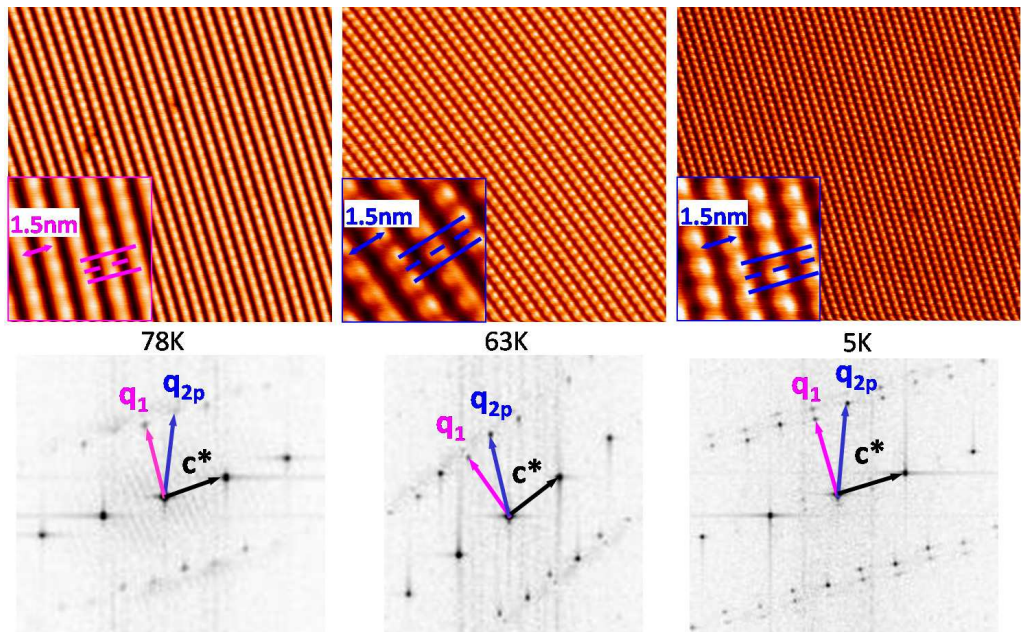}
\caption{\label{Q2Q1CDW_3T}(Color online) STM images of the \emph{in situ} cleaved $(\bm{b},\bm{c})$ surface of NbSe$_3$ measured at $T=$78, 63, and 5 K ($V_{bias}=-300$ mV, $I=100$ pA). Scanned areas: $60\times60$ to $80\times80$ nm$^2$. Insets: smaller portion of the image at larger scale. Below each image, its 2D Fourier transform shows the lattice Bragg spots (indicated by $\bm{c^*}$ normal to the chains) and the $\bm{q_1}$ and $\bm{q_2}$ superlattice spots. At 78 K the $\bm{q_1}$ CDW superlattice is clearly seen whereas the $\bm{q_2}$ superlattice spots are diffuse. At 63 K, i.e. 4 K above the bulk transition temperature, the $\bm{{q_{2p}}}$ CDW superlattice spots are already well defined, their amplitude being larger than the one of $\bm{q_1}$ spots. At 5 K the ratio of the $\bm{{q_{2p}}}$ to $\bm{q_1}$ amplitude is slightly larger than that at 63 K.}
\end{figure*}

NbSe$_3$, one of the most widely studied CDW system, was the first inorganic low-dimensional conductor to exhibit a sliding CDW state when an electric field above a threshold value is applied \cite{Monceau1976,Fleming1979,ECRYS}. It has a linear structure consisting of three pairs of metallic chains per unit cell running along the $\bm{b}$ axis, which are denoted as type I, II and III according to the strength of the chalcogen-chalcogen bond in the triangular basis of the chain. By STM we could unambiguously identify for the first time the three types of chains lying in the $(\bm{b},\bm{c})$ surface plane in the temperature range 5-140 K \cite{Brun2009a}. The surface CDW wave-vectors are $\bm{q_1} = 0.24$ $\bm{b^*}$ and $\bm{{q_{2p}}}= 0.26$ $\bm{b^*}$ $+$ $0.5$ $\bm{c^*}$, in excellent agreement with the bulk reported values \cite{Tsutsumi1977,Fleming1978,Smaalen1992} projected on the $(\bm{b},\bm{c})$ plane. A detailed analysis of the spatial distribution of the $\bm{q_1}$ and $\bm{q_2}$ CDWs on the various types of chain was performed and compared to bulk x-ray diffraction results \cite{Smaalen1992}. A long range new modulation defined by the wave-vector $ \bm{u} \simeq 2 \times (0.26 - 0.24)$ $\bm{b^*}$ was observed, resulting from the interaction of both CDWs being present on chains III \cite{Brun2009a}.

The present work was performed with an Omicron LT ultrahigh vacuum (UHV) STM system equipped with two UHV separated chambers. Well characterized NbSe$_3$ single crystals with typical dimensions of $0.01 \times 10 \times 0.05$ mm$^3$ were selected, cleaved \emph{in situ} at room temperature along the $(\bm{b},\bm{c})$ planes and cooled down to 5, 63 or 78 K. Samples were further thermalized at temperatures between 5 and 140 K. Our thermometer is a silicon diode in good thermal contact with the sample. Intermediate temperatures were stabilized by a feedback system controlling a heater element while the STM head is continuously cooled by the cryostat through copper braids. Experiments were conducted when the thermal drift became acceptable. Both mechanically sharpened Pt/Ir and electrochemically etched W tips were used leading to similar results. All the STM images shown in the following are measured with constant current.

Fig.~\ref{Q2Q1CDW_3T} shows three STM images measured at 78, 63 and 5 K, on large atomically flat terraces, with their 2D Fourier transform (2D FT). Strikingly at 78 K, almost 20 K above $T_{2b}$, one can easily detect the presence of diffuse $\bm{{q_{2p}}}$ superlattice spots, having a much lower amplitude than the $\bm{q_1}$ ones. At 63 K, i.e. 4 K above $T_{2b}$, the $\bm{{q_{2p}}}$ superlattice is well developed leading to sharp spots in the FT of the STM image, their amplitude being larger than the one of the $\bm{q_1}$ satellites. This indicates that the $\bm{q_2}$ CDW ordering occurs at higher temperature at the surface than in the bulk. It contrasts with the conventional behavior of the $\bm{q_1}$ CDW \cite{Brun2009a} for which we do not find a noticeable increment of the surface transition temperature over the bulk one.

\begin{figure*}
\includegraphics[width=17.5cm]{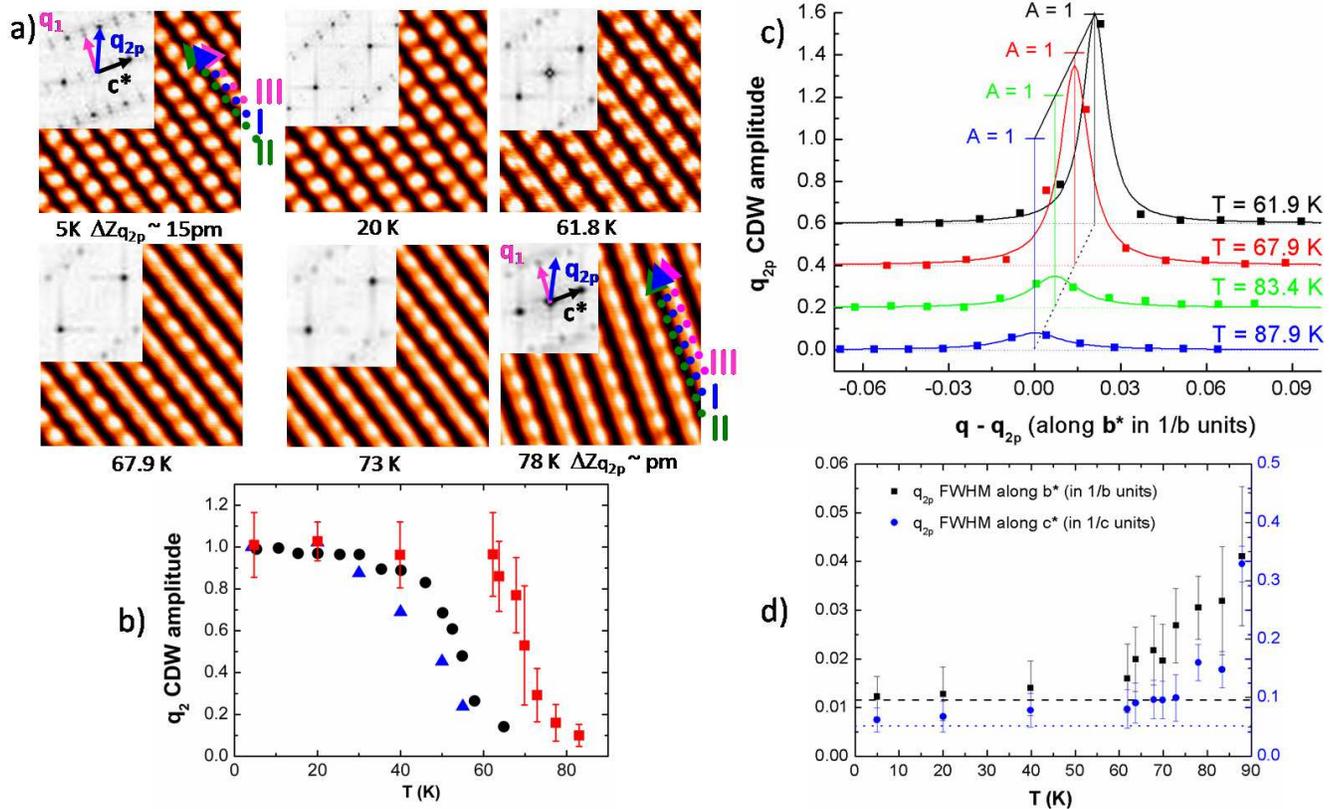}
\caption{\label{ImagesQ1Q2transition}(Color online) a) STM images ($10\times10$ nm$^2$) of the \emph{in situ} cleaved $(\bm{b},\bm{c})$ surface of NbSe$_3$ measured between 5 K and 140 K, showing the temperature evolution of the $\bm{q_{2p}}$ CDW ($V_{bias}=-300$ mV, $I=100$ pA). Insets: 2D Fourier transform (FT) of the corresponding $20\times20$ to $40\times40$ nm$^2$ image. b) squares: normalized amplitude of the $\bm{{q_{2p}}}$ CDW satellites measured by STM as a function of temperature (vertical bars: dispersion of the measurements); dots: x-ray diffraction results on bulk crystals \cite{Fleming1978}; triangles: double $\bm{q_2}$ energy gap probed by CDW-CDW tunneling \cite{O'Neill2006}. c) Profiles across the $\bm{{q_{2p}}}$ peak along the chain direction ($\bm{b^*}$) at different T - FT of the pair correlation function of the CDW amplitude over the STM image. d) Full widths at half-maximum (FWHM) along ($\bm{b^*}$) and perpendicular to the chains ($\bm{c^*}$) as function of T, extracted from FT of $30\times30$ nm$^2$ images - inverse correlation lengths. The dashed and dotted lines correspond to the smallest measurable FWHM along $\bm{b^*}$ and $\bm{c^*}$.}
\end{figure*}

The $\bm{q_2}$ CDW surface transition temperature surface ($T_{2s}$) was determined by analyzing the temperature dependence of the amplitude of the first-order $\bm{{q_{2p}}}$ satellites, extracted from 2D FT of STM images (areas: $20\times20$ to $40\times40$ nm$^2$) \cite{foot2}. A selection of the measured STM images is presented in Fig.~\ref{ImagesQ1Q2transition}a for $5<T<78$ K. As we have shown in \cite{Brun2009a}, only chains I and III are clearly visible in these tunneling conditions, while the location of chains II is shown for clarity, at 5 and 78 K. The corrugation of the $\bm{{q_{2p}}}$ CDW is about 15 pm at 5 K on chains I, has very weak variations up to 62 K, and decreases strongly between 62 K and 78 K. This is accompanied by a simultaneous diminution of the beating effect between $\bm{q_1}$ and $\bm{{q_{2p}}}$ on chains III (see \cite{Brun2009a}). At 78 K, the $\bm{{q_{2p}}}$ modulation nearly vanishes and the $\bm{q_1}$ wavefronts on chains III  becomes quasi-parallel and regular.

Quantitatively, the extracted $\bm{{q_{2p}}}$ Fourier amplitude as a function of temperature is shown in Fig.~\ref{ImagesQ1Q2transition}b, with x-ray diffraction results \cite{Fleming1978} and CDW-CDW tunneling data \cite{O'Neill2006}. A sharp transition is observed by STM in the range 70-75 K, much above $T_{2b}=59$ K \cite{Fleming1978}. From the results of \cite{O'Neill2006} the $T_{2s}$ of an unfree surface can be extrapolated to $\approx60$ K. Hence there is evidence for an "extraordinary transition" occurring at the $(\bm{b},\bm{c})$ surface of NbSe$_3$ for the $\bm{q_2}$ CDW, leading to CDW ordering at a temperature exceeding the bulk one by about 15-20\%. The temperature dependence is more abrupt in the vicinity of $T_{2s}$ at the surface than in the bulk. The dispersion of the experimental data reflects that at different positions along the surface the measured CDW corrugation is not identical, particularly in the vicinity of the transition around 70 K.

To explain the increase of $T_{2b}$ at the surface plane, it is appealing to consider the differences existing between phonons propagating at the surface and in the bulk. At a crystal surface, atom displacements normal to the surface are larger than those encountered in the bulk \cite{Desjonqu'eres1996}. Energies of the transverse phonon modes propagating in the surface plane are then smaller than those of the bulk phonons. The charge ordering occurring at the surface of NaCCOC has been recently explained in this way \cite{Brown2005}. Also the inelastic x-ray diffraction measurements on NbSe$_2$ have shown that the Kohn anomaly at $2\bm{k_F}$ is more pronounced at the surface and occurs at a lower energy than in the bulk underlying the changes in the phonon spectrum to explain the increase of $T_c$ at the surface \cite{Murphy2005}. We believe that a similar situation is likely to happen in NbSe$_3$. Above $T_{2b}$, when the $\bm{q_2}$ CDW fluctuations become 2D, these were shown to be transverse in the $(\bm{a},\bm{b})$ plane and not in the $(\bm{b},\bm{c})$ plane \cite{Rouziere1996}. This is consistent with x-ray refinements of the $\bm{q_2}$ superlattice modulation at 4 K, showing that larger lattice displacements occur along $\bm{a}$ than along $\bm{c}$ \cite{Smaalen1992}. Thus, the phonon modes involved in the $\bm{q_2}$ distortion possess a transverse component at the surface, supporting the idea of softer transverse phonon modes for the top NbSe$_3$ layer, which would lead to a CDW ordering at a higher critical temperature through an increased electron-phonon coupling. 

We get a deeper insight on the surface transition nature by measuring the in-plane correlation functions and extracting the inverse correlation lengths (see Figs.~\ref{ImagesQ1Q2transition}c and d). We observe a continuous evolution with $T$ showing that the system is in a 2D critical regime between 88 and 62 K. At 88 K, the correlations lengths $\xi_{b}$ (along $\bm{b}$) and $\xi_{c}$ (along $\bm{c}$) are equal to 25$b$ and 3$c$ respectively and increase to 70$b$ and 10$c$ at T$\approx$62K. Below 62 K both correlation lengths saturate, whereas the $\bm{{q_{2p}}}$ Fourier amplitude is stabilized according to Fig.~\ref{ImagesQ1Q2transition}b. Here, the 2D regime connects with the bulk pre-transitional region $T_{2b}=59$ K$<T<63$ K of 3D ordering \cite{Rouziere1996} - which will lead to a growing interaction with the bulk.

To analyze the experimental results further we should take into account some limitations regarding the length scale, a finite window in the STM experiment, as well as regarding the time scale, which is not instantaneous like e.g. in X-ray diffraction experiment. A window of size $L\times L$ with $L=30$ nm, comprises 85 lattice units - $20$ CDW periods - along $\bm{b}$ and $20$ lattice units - $10$ CDW periods - along $\bm{c}$. We see from Fig.~\ref{ImagesQ1Q2transition}d that $L=30$ nm determines the smallest measurable FWHM which are reached at $\approx62$ K. A useful scale to compare with is the expected correlation length at $T=70$ K due to the thermal fluctuations of an isolated chain, $\xi_{T}=\hbar v_{F}/(\pi k_BT)\approx10${\AA}$\approx3b$, with $v_{F}=3\cdot10^{7}cm/\sec$. This value is an order of magnitude smaller than the observed one, supporting the presence of the observed 2D regime. We infer from this analysis of the correlation lengths that at $62$ K appears a ghost transition because of i) the finite size of the observation window, ii) the beginning of 3D correlations in the bulk which suppresses the surface 2D fluctuations. 

The BKT transition refers to 2D systems like the He4, a superconductor or an incommensurate CDW, described by a complex order parameter. The BKT scheme contains two features; i) below $T_{BKT}$ the order parameter correlation falls off as a power law $\sim r^{-2\eta}$; ii) approaching $T_{BKT}$ the correlation length grows very fast, the nature of the transition being the confinement of vortex/anti-vortex pairs which may become unbound at $T>T_{BKT}$. In principle this description applies to the regime where the instantaneous local order parameter amplitude $A_{t,1D}$ (unlike the coherent component $A_{2D} \ll A_{t,1D}$ measured by the 2D FT) is already fixed, i.e. at low $T$ starting from a close vicinity of a mean-field (MF) transition at $T_{MF}$, which usually limits the observations. This region may be extended to higher $T$ in anisotropic systems like the CDWs, where the planes consist of chains. The 1D-2D-3D hierarchy \cite{Scalapino75} promotes the regime with phase-only fluctuations: the 1D regime coalesces directly into the BKT state \cite{Luther77}. In our case this facilitates the observation of the wide 2D fluctuating regime.

The lack of true long range order in space is ultimately related to its lack also in time. In the 1D regime, i.e. for $T>T_{BKT}$ (above $\approx90$ K), the estimated fluctuation time scale is orders of magnitude smaller than the typical STM acquisition time of 1 sec per line \cite{foot1}; hence the temporal fluctuations wash out the CDW from observations. For $T<T_{BKT}$, the tremendous slowing down of the fluctuations allows for the order observation (the local order time decay law $\sim t^{-\eta}$ is characterized by the very small $\eta\le 1/8$). This discussion leads us to a usually overlooked fundamental distinction between the slow STM measurement and the instantaneous one by X-rays. Grazing incidence x-ray diffraction would be suitable tools to compare with our conclusions by tracking the $\bm{q_2}$ transition at the surface and in the bulk. Finally we mention that some of our scans allow to visualize directly the vortices (dislocations in the CDW language) and their pairs (phase solitons) which strengthens the advocated BKT scenario.

In conclusion, high-resolution STM images of NbSe$_3$ show that the $\bm{q_2}$ CDW, unlike the $\bm{q_1}$ one, undergoes a sharp "extraordinary" transition at the surface. Our analysis of the correlation lengths shows at $\approx88$ K a transition from the 1D regime to the 2D BKT state. This seems to be the first real space observation of the BKT regime. Below $\approx63$ K, 2D correlations are stabilized by the critical increase of 3D correlations in the bulk. When probing the fluctuations of the local CDW order by STM, our analysis emphasizes that the time decay of the local order dominates over the space one. This conclusion may have impact on other STM studies in strongly fluctuating systems like stripes in doped oxides.

The high-quality NbSe$_3$ samples were synthesized by H. Berger and F. L\'evy. We thank J.-C. Girard, E. Machado and E. Canadell for many helpful discussions and C. David for technical assistance. This work was partly supported by 
the ANR grant BLAN07-03-192276.

\bibliography{biblioNbSe3}

\begin{thebibliography}{21}
\bibitem[{foo({\natexlab{a}})}]{foot2}
\bibinfo{note}{To avoid accounting for the temperature dependence of the
  piezo-elements in the measured Fourier amplitude, the latter were normalized
  to the $\bm{q_1}$ amplitude.}
\bibitem[{foo({\natexlab{b}})}]{foot1}
\bibinfo{note}{The CDW friction coefficient is inferred from the known
  nonlinear conductivity.}
\end{thebibliography}

\end{document}